\documentclass[floatfix,onecolumn,aps,prd,amsmath,amssymb]{revtex4}
\usepackage[latin1]{inputenc}
\usepackage[dvips]{graphicx}

\usepackage{dcolumn}
\usepackage{bm}
\usepackage{color}
\usepackage{url}
\usepackage[colorlinks=true ,citecolor=blue]{hyperref}
\usepackage{amsmath,amssymb}

\newcommand{\PC}[1]{\ensuremath{\left(#1\right)}}

\newcommand{\n}{\noindent}

\begin{document}

\title{Quantum-electrodynamical approach to the exciton spectrum in Transition-Metal Dichalcogenides}

\author{E. C. Marino$^1$, Leandro O. Nascimento$^{1,2,3}$, Van S\'ergio Alves$^{1,4}$, N. Menezes$^2$, and C. Morais Smith$^2$}
\affiliation{$^1$Instituto de F\'\i sica, Universidade Federal do Rio de Janeiro, C.P.68528, Rio de Janeiro RJ, 21941-972, Brazil \\
$^2$Institute for Theoretical Physics, Centre for Extreme Matter and Emergent Phenomena, Utrecht University, Princetonplein 5, 3584CC Utrecht, The Netherlands \\
$^3$ International Institute of Physics, Campus Universit\'ario-Lagoa Nova-59078-970-CP: 1613 - Natal/RN-Brazil \\
$^4$Faculdade de F\'\i sica, Universidade Federal do Par\'a, Av.~Augusto Correa 01, 66075-110, Bel\'em, Par\'a, Brazil. }

\date{\today}

\maketitle  
{\bf Abstract}

{\bf  Manipulation of intrinsic electron degrees of freedom, such as charge and spin, gives rise to electronics and spintronics, respectively. Electrons in monolayer materials with a honeycomb lattice structure, such as the Transition-Metal Dichalcogenides (TMD's), can be distinguished according to the region (valley) of the Brillouin zone to which they belong. Valleytronics, the manipulation of this electron's property, is expected to set up a new era in the realm of electronic devices. In this work, we accurately determine the energy spectrum and lifetimes of exciton (electron-hole) bound-states for different TMD materials, namely WSe$_2$, WS$_2$ and MoS$_2$. For all of them, we obtain a splitting of the order of 170 meV between the exciton energies from different valleys, corresponding to an effective Zeeman magnetic field of 1400 T. 
Our approach, which employs quantum-field theory (QFT) techniques based on the Bethe-Salpeter equation and the Schwinger-Dyson formalism, takes into account the full electromagnetic interaction among the electrons. 
The valley selection mechanism operates through the dynamical breakdown of the time-reversal (TR) symmetry, which originally interconnects the two valleys. This symmetry is spontaneously broken whenever the full electromagnetic interaction vertex is used to probe the response of the system to an external field.} 

\bigskip

{\bf Introduction}

Understanding the properties of exciton bound states in solids is crucial for applications  in optoelectronics, such as photovoltaics \cite{furchi}, phototransistors \cite{wang,Sanchez}, and light-emitting diodes \cite{Zhang,Cheng}, to cite just a few. The recent synthesis of monolayer XM$_2$ TMD's, where X = Mo, W, Sn and M = S, Se, Te, has revealed materials that might be ideal for valleytronics because of their large, direct, and non-centrosymmetric gap. Indeed, due to a strong electrostatic interaction, unusually large exciton binding energies of a few hundreds of meV emerge inside a large electronic bandgap, of a few eV \cite{review1,review2}. 

A fundamental issue for valley manipulation is to understand the role played by each valley in the process of exciton formation after the photoexcitation of an electron-hole pair. Subsequently, valley manipulation methods can be devised, by taking advantage of such valley-dependent exciton properties. Resonant excitation of electron-hole pairs with linearly polarized light, for instance,  generates coherent linear combinations of excitons belonging to the $K$ and $K'$ valleys in TMD materials. Valley manipulation has, then, been achieved by rotating the relative phase of such coherent exciton states in monolayer TMD's through an
explicit breaking of TR-symmetry. This happens either by an applied magnetic field \cite{2,2a,2b,2c}, of the order of 10 T, or by the use of circularly polarized light \cite{3,3a}, which couples selectively to each valley, and has enabled phase rotations on the scale of femtoseconds \cite{4,1}. The excitonic spectrum of TMDs is typically determined experimentally by using photoluminescence excitation spectroscopy (PLE)  \cite{PLE} and time-resolved mid-infrared spectroscopy \cite{midIR}. Theoretical studies are mostly based on elaborated first-principle calculations that include spin-orbit coupling to obtain a realistic band structure. The Coulomb interaction alone is unable to explain the experimental data for the exciton binding energies \cite{WS2expp}, hence the Keldysh potential \cite{Keldyshp,a} has been frequently used to describe the excitons in TMD's \cite{np}. The use of a static potential, however, rules out some important many-body effects such as dynamical TR-symmetry breaking.

Here, we propose an alternative approach, which starts from a gaped relativistic band structure (Dirac with a gap) originating from a simplified tight-binding expansion around the $K$ and $K'$ valleys. It lacks, however, corrections such as the spin-orbit coupling. One then includes the {\it full dynamical electromagnetic interaction} containing the complete sequence of interaction terms. A Foldy-Wouthuysen expansion of our model shows that in the {\it semiclassical limit}, it yields a parabolic band with spin-orbit coupling, in addition to other terms, such as an electrostatic potential and a Darwin term. Moreover, we show that the Keldysh potential emerges as the {\it static limit} of our theory for distances larger than a characteristic length scale, proportional to the inverse of the gap, whereas for small distances one obtains the usual Coulomb interaction \cite{footnote}. Our approach allows one to calculate the excitonic spectrum, as well as its lifetime, and yields a very good agreement with experiments on WS$_2$ \cite{WS2expp}, WSe$_2$ \cite{WSE2expp}, and MoS$_2$ \cite{PLE}. It also shows that quantum-field theory techniques not only provide an illuminating and elegant perspective of the problem, since they allow to retrieve in the semiclassical and static limit the usual terms used in more conventional approaches (parabolic-band structure, spin-orbit coupling, and Keldysh potential), but they reveal the appearance of anomalies and symmetry breaking, which are truly quantum-many-body features. We show here that this is indeed the case for TMDs, and propose an experiment to verify our theoretical predictions. \\

{\bf Model}

The tight-binding energy spectrum of TMDs reveals two bands separated by a gap at the $K$ and $K'$ points of the Brillouin zone, reminiscent of the spectrum of a massive Dirac Hamiltonian. Low-lying excitations are thus described by a gaped Dirac theory. In order to go beyond the tight-binding approximation, one must introduce the electromagnetic interaction among these Dirac electrons. However, there is a mismatch between the dimensionality of the electrons/holes kinematics, which are constrained to the two-dimensional (2D) plane, and the 3D dynamics of the photons mediating their interaction. Hence, a projected formalism, the so-called Pseudo Quantum Electrodynamics (PQED) \cite{marinop}, is the appropriate tool to study the effect of the full  interactions in this kind of systems \cite{PRX}. In this approach, one starts with Maxwell QED in 3+1D and projects the dynamics of the photons into the 2D plane, thus obtaining an effective Lagrangian that is non-local because parts of the system have been integrated out. The name PQED stems from the fact that a pseudo-differential operator, proportional to the inverse square root of momentum, arises after the projection \cite{marinop}. Despite the non-locality, the model is unitary and dual to Maxwell QED in 2+1D. Indeed, the Green's function (propagator) for PQED in momentum space behaves like the one for Maxwell QED in real space and vice-versa \cite{unit}. 

The Lagrangian model describing the system is given by \cite{marinop}
\begin{equation}
{\cal L}= \frac{1}{2} F_{\mu \nu}\left(\frac{1}{\sqrt{-\Box}}\right) F^{\mu\nu}+\bar\psi_a \Big(i\partial\!\!\!/-M^a\Big)\psi_a+j^{\mu}A_{\mu}\, ,
\label{action}
\end{equation}
where $i\partial\!\!\!/ =i\gamma^0\partial_0+i\,v_F\gamma^i\partial_i$ and $j^\mu=e\,\bar\psi\gamma^\mu\psi=e\,(\bar\psi\gamma^0\psi,v_F \,\bar\psi\gamma^i\psi)$.
Here, $\psi_a=(\psi_A,\psi_B)_a$ is a two-component Dirac field, corresponding to the inequivalent $A$ and $B$ sublattice sites of the honeycomb lattice, $a = K\uparrow, K'\uparrow, K\downarrow, K'\downarrow $ is a flavor index accounting for the spin and valley internal degrees of freedom, $\bar\psi_a =\psi_a^\dagger\gamma^0$, and $F_{\mu \nu}$ is the usual field-intensity tensor of the U(1) gauge field $A_\mu$, which intermediates the electromagnetic interaction in 2D (pseudo electromagnetic field), with $\gamma^\mu$ rank-2 Dirac matrices. The coupling constant $e^2= 4\pi\alpha$ is conveniently written in terms of $\alpha$, the fine-structure constant in natural units. The mass term has different signs for each valley, $M^a=\xi \Delta$, where $\Delta>0$ and $\xi=\pm 1$, respectively, for valley $K$ and $K'$, i.e. $M^a=(M^{K}, M^{K'})_\sigma = (\Delta,-\Delta)_\sigma$, where $\sigma = \uparrow, \downarrow$ \cite{Ezawap,Ezawap2,Ezawap3,wsilp}. It arises due to the different atoms forming the TMDs (Mo and S, e.g.), which lead to a staggered chemical potential in the tight-binding description, and ultimately imply into masses with opposite sign in the Dirac description around each valley.  The opposite signs of the masses make the Lagrangian invariant under time-reversal symmetry. In the Sup. Mat. (Sec. 1-3), we provide a detailed derivation of the model,  starting from the usual tight-binding approach. In addition, we show its time-reversal invariance, and recall some important properties of the Dirac field, which describes both electrons and holes, as well as their interactions. \\

{\bf Procedure}
 
In a QFT description, excitons appear as poles in the Fourier transform of the two-particle Green's function $G(x;y)= \langle 0|T \psi(x) \psi^\dagger(x) \psi(y) \psi^\dagger(y)|0\rangle$, namely 
$G(p)$. This is related to the exact interacting electron/hole propagator $S(p)$ through the Bethe-Salpeter equation \cite{izp},
\begin{equation}
G(p)=\frac{1}{\Big[S^{-1}(p) \Big]^2 - \Gamma(p)}, 
\label{bs}
\end{equation}
or, equivalently,
\begin{equation}
G^{-1}(p)=G_0^{-1}(p) - \Gamma(p), 
\label{bs2}
\end{equation}
where $G_0(p)= S^2(p)$ is the free exciton propagator, which is nothing but the square of the dressed electron/hole propagator, $S(p)$. This, by its turn, satisfies the Schwinger-Dyson equation
\begin{equation}
S^{-1}(p)= S_0^{-1}(p) - \Sigma(p) = p\!\!\!/ - M_{a} - \Sigma(p),
\label{bs1}
\end{equation}
where $S_0(p) = [p\!\!\!/ - M_{a}]^{-1}$ is the free electron propagator, with $M_a$ denoting the bare gap, and
$\Sigma(p)$ is the electron/hole self-energy. The structure of Eq.~(\ref{bs}) reveals  that $\Gamma(p)$, the exciton interaction kernel, acts as the exciton self-energy (see the analogy by looking at Eq.~(\ref{bs1}) and notice that the first term in Eq.~(\ref{bs2}) describes the electron and hole interactions, whereas the second describes the exciton interactions. 
The poles of $G(p)$ occur at $G^{-1}(p) = 0$, namely $p\!\!\!/= M_a + \Sigma(p) \pm  \Gamma(p)^{1/2}$.

In order to determine the exciton energy spectrum, we will calculate the exciton poles by neglecting the exciton-exciton interaction kernel $\Gamma(p)$. In this approximation, $p\!\!\!/= M_a + \Sigma(p) \equiv M_a^R$. Since we are neglecting the renormalization of the exciton binding energy due to the exciton-exciton interactions, the exciton poles are given in terms of the electron/hole self-energy alone, which now has to be determined self-consistently and evaluated at $p =  M_a^R$. This approximation is justified by the fact that the excitons are neutral objects, and hence the exciton-exciton interaction should be much weaker (of the order of ten times smaller) than the corresponding electron-hole interaction (see Sec. 4 in the Sup. Mat. for details of the calculation and for an estimate of this interaction). 
Therefore, the exciton eigenenergies are given by
\begin{equation}
\varepsilon_a^{\pm}=\pm |M_a^R|,
\label{ee}
\end{equation}
where the plus/minus sign arises because the propagator in the denominator of Eq. (\ref{bs}) is squared. 

\begin{figure}[hbt]
\centering
\includegraphics[scale=0.27]{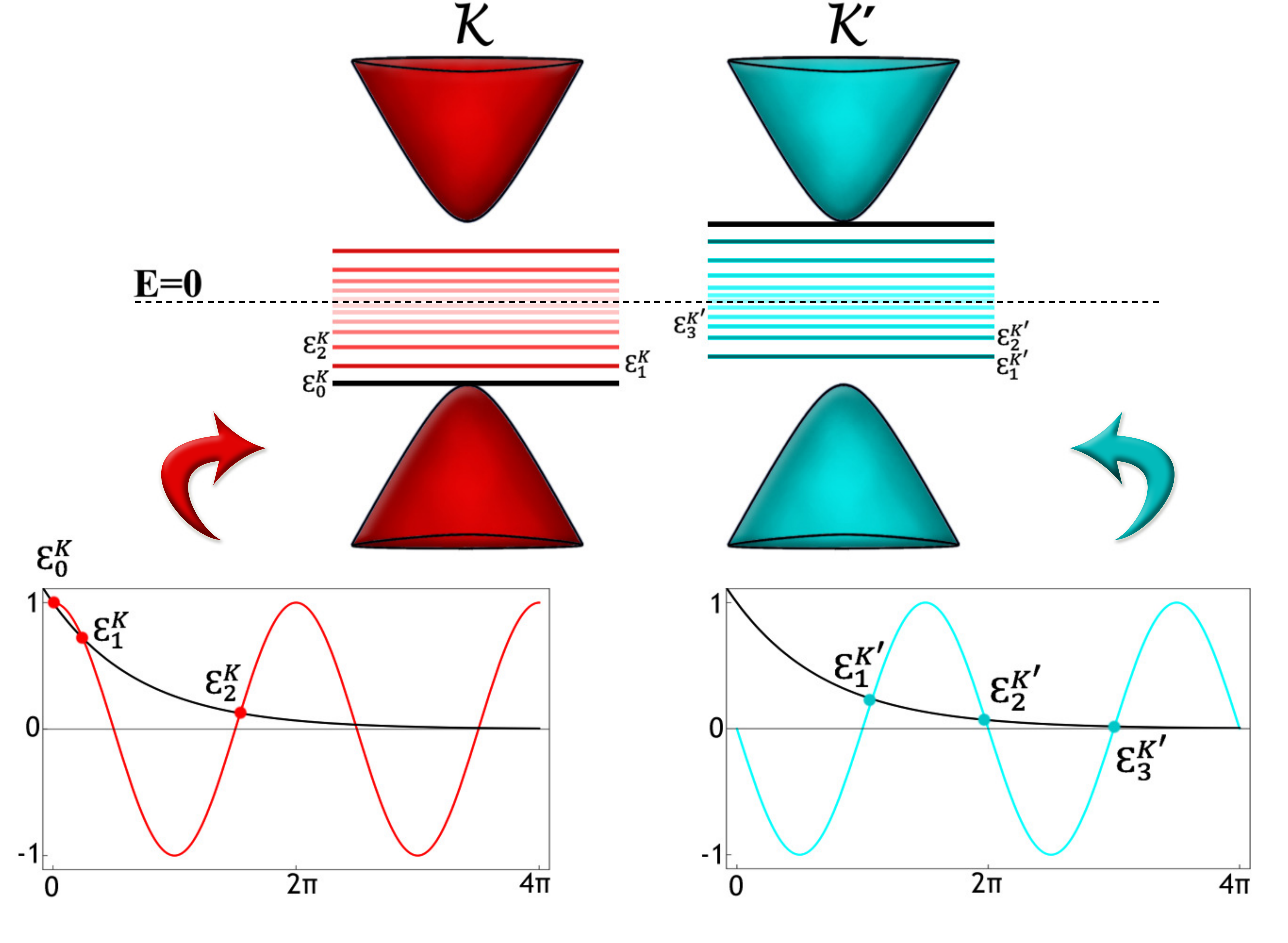}
\caption{Sketch of the midgap states in the valleys $K$ and $K'$. The additional unpaired state at the top of the valence band in valley $K$ (bottom of the conduction band in the valley $K'$) occurs as a consequence of the parity anomaly. The excitonic states are solutions of the transcendental equations represented in the bottom of the figure.} \label{Fig1}
\end{figure}

We then proceed to determine $ \Sigma_a(p) $
by solving the non-perturbative Schwinger-Dyson equation for each flavor $a$. 
The Schwinger-Dyson equation leads to a differential equation for $\Sigma_a(p)$, the solution of which is such that at momentum $\textbf{p}=0$ there are two series of energy eigenvalues, namely $\varepsilon(X_n)=\Lambda e^{- X_n/\gamma}$  and  $\varepsilon(Y_n)=\Lambda e^{- Y_n/\gamma}$, $n=1,2,3,...$, where 
$X_n$ and $Y_n$ are, respectively, the solutions of the transcendental equations
\begin{eqnarray}
 e^{-3z/2\gamma}=\cos z\ \ \ ;\ \ \  e^{-3z/2\gamma}=-\sin z, 
\label{2}
\end{eqnarray}
which correspond, respectively, to each of the two valleys mass signs.
In the above expressions, $\Lambda = \hbar v_F /a$ is an energy-momentum cutoff related to the Fermi velocity $v_F$ and the lattice constant $a$, and 
\begin{equation}
 \gamma = \frac{1}{2}\sqrt{\frac{16 \alpha }{ \pi(\pi \alpha - 2)} - 1}
\label{gamma}
\end{equation}  
is a function of the fine-structure constant  $\alpha = {\rm e}^2/4 \pi \epsilon_0 \epsilon v_F$, which  is supposed to be larger than a critical value $\alpha_c =2/\pi \approx 0.63 $ (see Sec. 5 of the Sup. Mat.). This fitting parameter is connected to the dielectric constant of the material/substrate $\epsilon$. 

Now, we have for the valley $K$ that the renormalized mass $M_a^R= \Delta + \Sigma(M_a^R)$, whereas for the valley $K'$, $M_a^R=- \Delta + \Sigma(M_a^R)$. The energies of such bound states at zero momentum $\textbf{p}=0$ are given, respectively, by
\begin{eqnarray} \label{MpmK}
\varepsilon_n^+(Y_n)&=& \Lambda e^{-Y_n/\gamma}, \ \ \ \ \ \  \varepsilon_n^-(X_n)=- \Lambda e^{-X_n/\gamma}, \ \ \ \  {\rm valley} K \\
\varepsilon_n^+(X_n)&=& \Lambda e^{-X_n/\gamma},  \ \ \ \ \ \ \varepsilon_n^-(Y_n)= - \Lambda e^{-Y_n/\gamma} \ \ \ \  {\rm valley} K'.
\label{MpKprime} 
\end{eqnarray}
A sketch of the exciton energies obtained from Eqs. (\ref{MpmK}) and (\ref{MpKprime}) is presented in Fig.~\ref{Fig1}. 

The real part of the self-energy provides the renormalized energies, whereas the imaginary part is related to the lifetime of these excitations. 
However, this would be the intrinsic lifetime, corresponding to a direct electron-hole annihilation into a photon. Now, since the photon momentum is very small because $|\textbf{k}|=\omega/c$, it follows that for most of the excitons, the initial momentum cannot be matched to guarantee momentum conservation during the decay. As a result, the exciton decay should occur through a more complicated process, in which the exciton lifetime will be considerably enlarged, depending on the temperature \cite{excitlifep}. We show in the Sup. Mat. that an exciton lifetime enlargement of about two orders of magnitude occurs at $T\simeq 7$K. \\

{\bf Results}

\begin{figure}[tbh]
\centering
\includegraphics[scale=1.0]{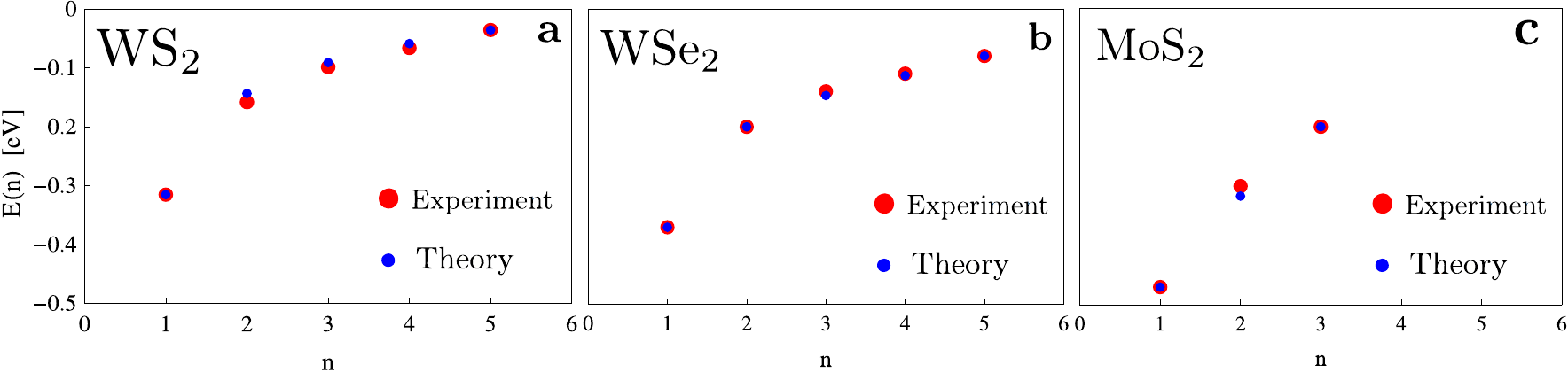}
\caption{\textbf{a} Comparison between the experimental data of the exciton energy measured in Ref.~\cite{WS2expp} for monolayer WS$_2$ and our theoretical exciton energy from Eqs.~(\ref{MpmK}) and (\ref{MpKprime}). \textbf{b} Comparison for monolayer WSe$_2$~\cite{WSE2expp}. \textbf{c} Comparison for monolayer MoS$_2$~\cite{PLE}. The red dots represent the experimental data, while the blue dots are our theoretical results. By fitting the first and last energy values, we find $\Lambda=0.4066$ eV and $\gamma=3.2251$ (or correspondingly, $\alpha=0.66$)  for WS$_2$, whereas for WSe$_2$ we obtain $\Lambda=0.42$ eV, $\gamma=4.68$, and $\alpha=0.65$,  and for MoS$_2$ we find $\Lambda=0.49$ eV, $\gamma=8.48$, and $\alpha=0.64$. } \label{fig2}
\end{figure}

A comparison of our results for the exciton-energy spectrum, contained in  Eqs. (\ref{MpmK}) and (\ref{MpKprime}), applied to three monolayer TMD's, namely, WS$_2$, WSe$_2$, and MoS$_2$, is shown in Fig.~\ref{fig2}.
 In the Sup. Mat. (Sec. 6), we provide a detailed explanation on how we have used our theoretical findings to evaluate the electron-hole binding energies and lifetimes in different materials.

\textit{A -- Monolayer of} WS$_2$: We extract the energy splitting between the exciton-energy level and the bare-energy gap from the five exciton states measured for WS$_2$ (see Fig.~3 in Ref.~\cite{WS2expp}), and find $\varepsilon^{{\rm exp}}_{b,n}=\{-0.3155, -0.1584,-0.0989,-0.0664,-0.0359\}$ eV. By comparing our expressions with two of the above experimental data (first and last points), we fix the two unknown parameters in the theory, namely $\Lambda=0.4066$ eV and $\gamma=3.2251$. Our theoretical expressions  Eqs. (\ref{MpmK}) and (\ref{MpKprime}) then yield $\varepsilon^{{\rm Theo.}}_{b,n}=\{-0.3155,-0.1437,-0.0912,-0.0589,-0.0359\}$ eV, see Fig.~\ref{fig2}a for a comparison. Notice that the three intermediate values were determined without any fitting parameter, and the same value of $\alpha=0.66$ (extracted from $\gamma$) has been used for all points.  

\textit{B -- Monolayer of} WSe$_2$: Using the same procedure as before, we extract the five exciton states for WSe$_2$ (see Fig.~4 in Ref.~\cite{WSE2expp}) from the experimental data, $\varepsilon^{{\rm exp}}_{b,n}=\{-0.37, -0.20,-0.14,-0.11,-0.08\}$ eV. The theoretical values are $\varepsilon^{{\rm Theo.}}_{b,n}=\{-0.37,-0.20,-0.15,-0.11,-0.08\}$ eV, which were obtained for $\Lambda=0.42$ eV and $\gamma=4.68$, corresponding to $\alpha=0.65$. The comparison between theory and experiment can be promptly visualized in Fig~\ref{fig2}b. 

\textit{C -- Monolayer of} MoS$_2$: Repeating again the same steps as before, we extract the three exciton energies $\varepsilon^{{\rm exp}}_{b,n}=\{-0.47, -0.30,-0.20\}$ eV from Ref.~\cite{PLE}. Our theoretical results are now $\varepsilon^{{\rm Theo.}}_{b,n}=\{-0.47,-0.32,-0.20\}$ eV, after using two experimental points to determine $\Lambda=0.5$ eV and $\gamma=8.48$, which corresponds to $\alpha=0.64$. The results are depicted in Fig~\ref{fig2}c.

In the above sequences of exciton energies, the values alternate between the two valleys, corresponding to $X_n$ and $Y_n$ solutions of each of the transcendental equations depicted in Fig. 1. Hence, we obtain intervalley energy splittings of up to 170 meV, which correspond to an effective magnetic field of 1400 T.

Interestingly, although the value of $\gamma$ changes for the different materials, the value of $\alpha$ is nearly constant for all compounds, $\alpha=0.66 - 0.64$, and is in excellent agreement with the value otained in Ref.~\cite{Markalphap}, namely $\alpha=0.7$. 

\textit{D -- Exciton Lifetimes}: 
The imaginary part of our solution for the inverse self-energy yields an intrinsic lifetime $\tau_0\simeq 2h[\cos X_n]^{1/3}/[\Lambda \sin X_n]$ for the most stable excitons. According to our approximation, the result does not incorporate the corrections coming from exciton-exciton interactions. Inserting the above values of $\Lambda$ and $X_1$ (see Sec. 7 in Sup. Mat.), we find the intrinsic lifetimes $\tau= 48.34$ fs for  MoS$_2$; $\tau= 32.4$ fs for  WSe$_2$ and $\tau= 24.7$ fs for  WS$_2$. At $T=7$ K, this lifetime is enlarged by a factor of about two orders of magnitude \cite{excitlifep}, thus yielding values of a few picoseconds for the effective exciton lifetimes:  $\tau_{eff}= 4.2$ ps for  MoS$_2$; $\tau_{eff}= 2.8 $ ps for  WSe$_2$ and $\tau_{eff}= 2.16$ ps for  WS$_2$. Typical lifetime values measured in experiments made at such temperature are in the range $1-10$ ps \cite{excitlifep}.

The theoretically obtained exciton energies are in excellent agreement with the available experimental data for the excitonic spectrum \cite{PLE, WS2expp, excitlifep}, and the same holds for the predicted exciton-lifetime values \cite{excitlifep}. Our results also enable to understand why just a few excitonic bound states can be observed, as their lifetimes quickly vanish when their energy is increased. 
A second interesting feature of our general formalism, is that it reduces to the more conventional two-body Keldysh interaction potential in the particular case of {\it static} electrons and holes, thereby establishing the link between this dynamical QFT method and the more commonly used Keldysh approach (see Sec. 8 in Sup. Mat. for a detailed derivation).  \\

{\bf Discussion}

In this work, we have shown that a breakdown of the time-reversal symmetry and, by extension of the valley degeneracy, is produced dynamically through the generation of midgap exciton states, in such a way that the excitonic spectrum of each valley is different. The splitting energy is of the order of 170 meV, corresponding to a Zeeman effect of an effective field of $1400$ T. Furthermore, for a given value of the chemical potential (for instance $\mu =0$), the total number of dynamically generated states is different for each of the valleys. The existence of an asymmetric $n = 0$ state in only one of the valleys implies, via the bulk-boundary correspondence, the presence of an associated quantized  edge current, which characterizes the spontaneous onset of an emergent quantum Hall effect \cite{PRL}. This result is here confirmed by the occurrence of a topological quantum anomaly in two dimensions, analogous to the axial anomaly that exists in massless Dirac-fermion systems in three spatial dimensions \cite{PRL}. The phenomenon is reminiscent of the parity anomaly predicted to arise in two-dimensional massive Dirac systems subjected to a perpendincular magnetic field \cite{Semenoff}, with the difference that instead of  an imbalance in the number of Landau levels,  there is an imbalance in the number and energy values of midgap states. The $n = 0$ is {\it not} an excitonic bound state, it has an infinite lifetime, and corresponds thus to a current-carrying state. 

Although we predict different energy states for each valley, and circularly polarized light should in principle be able to select only one of the valley contributions, no different spectrum is observed experimentally for light polarized in one direction or another. This is however, not in contradiction with our results. Indeed, ascribing the mass sign for each valley is a completely arbitrary procedure. As it happens in a magnet, e.g., the system should form domains where the signs of the masses of both $K$ and $K'$ valleys, are alternatively (+-) or (-+). This implies that the exciton spectrum of each valley alternates between the solutions of each of the two transcendental equations in Fig. 1 in each domain, and both sets of solutions are always observed. A possibility to disentangle the two sets of solutions could arise in very small domains, as recently observed to form in chemical-vapor deposited WS$_2$ samples \cite{italiano}. Indeed, different domains, of characteristic size of the order of $10 \mu m$ have been observed recently \cite{italiano}. PLE studies reveal that the exciton peaks, despite having almost the same energy in the different domains, yet present quite different intensities \cite{italiano}. It would be extremely interesting to perform the experiment reported in Ref.  \cite{italiano}, but now using circularly polarized light. The occurrence of substantial exciton energy shifts in the different domains would configure the experimental detection of the phenomenon discussed here. 

One could possibly employ our procedure to study bi-excitons in monolayer TMD's. For this, however, we should not neglect the Bethe-Salpeter kernel $\Gamma(p)$, which precisely conveys the exciton-exciton interaction leading to the formation of bi-excitons. Recent progress in sample production, achieved upon encapsulating TMDs in hBN, has shown that much sharper excitonic detection is possible, by considerably reducing the linewidth. The generation of ultra-clean flakes, allied to the recent progress in spectroscopic techniques, holds promises to the observation of these fascinating and unexpected valley-selective excitons, rooted in the realm of quantum many-body dynamic interactions.

\acknowledgments

This work was supported in part by CNPq (Brazil), CAPES (Brazil), FAPERJ (Brazil), and by the Brazilian project Science Without Borders. The work by C.M.S. is part of the D-ITP consortium, a program of the Netherlands Organization for Scientific Research (NWO) that is funded by the Dutch Ministry of Education, Culture and Science. V.S.A. acknowledges the Institute for Theoretical Physics of Utrecht University for the kind hospitality and CNPq for financial support. Leandro O. Nascimento thanks the Ministry of Science, Technology and Innovation of Brazil and the Ministry of Education and Culture of Brazil.
We are grateful to Tony Heinz for fruitful discussions.\\

\bigskip
{\bf Competing Interests}

The authors declare that they have no competing financial interests.

\vfill\eject

\begin{center}
\textbf{\large Supplementary Information:}\\
\textbf{\large Quantum-electrodynamical approach to the exciton spectrum in Transition-Metal Dichalcogenides}
\end{center}
\vspace{1.5cm}

{\bf
 1. From Tight-Binding to the Dirac Description } \\

TMD's have a honeycomb lattice structure with different atoms occupying sublattices $A$ and $B$.
In addition to the usual inter-lattice hopping terms, therefore, the tight-binding Hamiltonian will have a staggered chemical potential on the sublattices. Introducing the field
 $\Psi^\dagger(\textbf{k},\sigma)=(c^\dagger_{A}(\textbf{k},\sigma)\ c^\dagger_{B}(\textbf{k},\sigma))$, where 
$c^\dagger_{A,B}(\textbf{k},\sigma)$
is the creation operator of an electron with momentum $\textbf{k}$ and  spin components $\sigma=\uparrow, \downarrow$ in sublattices $A$ and $B$,
we may express the tight-binding Hamiltonian as 
\begin{eqnarray}
H_{TB} = \sum_{\textbf{k},\sigma}\Psi^\dagger(\textbf{k},\sigma)
 \left (
\begin{array}{ll}
M & \phi \\
          \phi^* & -M
\end{array}
\right)
\Psi(\textbf{k},\sigma),
\label{1330}
\end{eqnarray}
where
$\phi (\textbf{k})$ is given by 
\begin{eqnarray}
\phi (\textbf{k})= -\tilde t  \sum_{i=1,2,3} 
e^{i \textbf{k}\cdot \textbf{d}_i}, 
\label{1281}
\end{eqnarray}
with $\tilde t$ denoting the nearest-neighbor hopping, and the nearest-neighbor vectors $ \textbf{d}_i$ read 
\begin{eqnarray}
\textbf{d}_1 = \frac{a}{\sqrt{3}}\hat{y},  \ \ \   \textbf{d}_2 = -\frac{a}{\sqrt{3}}\left(\frac{\sqrt{3}}{2}\hat{x}+\frac{1}{2} \hat{y}\right), \ \ \ \textbf{d}_3 = \frac{a}{\sqrt{3}}\left(\frac{\sqrt{3}}{2}\hat{x}-\frac{1}{2} \hat{y}\right).
\label{1276}
\end{eqnarray}
Notice that the  $M$-term, being proportional to a $\sigma_z$ matrix in the $(A,B)$ space, describes the staggered energy arising  from the local atomic asymmetry between the two sublattices in TMD's. 

For the sake of completeness, we now show how to obtain the Dirac Hamiltonian from the tight-binding model. We follow Ref.\cite{em}, and add some details when necessary. Let us start by rewriting the Hamiltonian as 
\begin{eqnarray}
H_{TB} &=& \sum_{\textbf{k},\sigma}\Psi^\dagger(\textbf{k},\sigma)
h_{TB} \Psi(\textbf{k},\sigma),
\nonumber \\
\nonumber \\
h_{TB} &=&-\tilde{t} \left[\sum_i\cos(\textbf{k}\cdot \textbf{d}_i )\sigma_x + \sum_i\sin (\textbf{k}\cdot \textbf{d}_i)\sigma_y
\right ]
+M \sigma_z, 
\label{1332}
\end{eqnarray}
where the $\sigma_i$'s, for $i = x,y,z$ are Pauli matrices with entries in
 the $(A,B)$ space.
 
Next, we perform an expansion around the points 
$$
\textbf{K}=\frac{4\pi}{3a} \hat{x},  \quad   \textbf{K}'=-\frac{4\pi}{3a} \hat{x},
$$ 
known as valleys, and
defined in such a way that $\phi(\textbf{K})= \phi(\textbf{K}')=0$.
Writing $\textbf{k}=\textbf{Q}+\textbf{p}$, for  $\textbf{Q}=\textbf{K},\textbf{K}'$, respectively, we obtain
\begin{eqnarray}
H_{K} &=& \sum_{\textbf{k},\sigma}\Psi^\dagger_{K}(\textbf{k},\sigma)
h_{K} \Psi_{K}(\textbf{k},\sigma),
\nonumber \\
\nonumber \\
H_{K'} &=& \sum_{\textbf{k},\sigma}\Psi^\dagger_{K'}(\textbf{k},\sigma)
h_{K'} \Psi_{K'}(\textbf{k},\sigma),
\label{1285x}
\end{eqnarray}
where
\begin{eqnarray}
h_{K} &=&\frac{\sqrt{3}\tilde{t}a}{2} \left[p_x\sigma_x + p_y \sigma_y + M \sigma_z\right ] + O(p^2),
\nonumber \\
\nonumber \\
h_{K'} &=&\frac{\sqrt{3}\tilde{t}a}{2} \left[- p_x\sigma_x + p_y \sigma_y + M \sigma_z\right ] + O(p^2),
\label{1333}
\end{eqnarray}
and
$$
 \Psi_{Q}(\textbf{p},\sigma) =  \Psi_{Q}(\textbf{Q}+\textbf{p},\sigma)
$$
for  $\textbf{Q}=\textbf{K},\textbf{K}'$, and $|\textbf{p}|\ll |\textbf{Q}|$.

Then, performing the canonical transformation 
\begin{eqnarray}
 &\ &\Psi_{K'}(\textbf{k},\sigma)\longrightarrow\exp\Big\{i\frac{\pi}{2}\sigma_z\Big\}\sigma_x \Psi_{K'}(\textbf{k},\sigma), 
\nonumber \\
 &\ &\Psi_{K'}^\dagger(\textbf{k},\sigma)\longrightarrow\Psi_{K'}^\dagger(\textbf{k},\sigma)\sigma_x \exp\Big\{-i\frac{\pi}{2}\sigma_z\Big\},
\label{1287x} 
\end{eqnarray}
in $\Psi_{K',\sigma}(\textbf{k},\sigma)$, we see that the mass term transforms as
\begin{eqnarray}
&\ &M\Psi^\dagger_{K',\sigma} \sigma_z \Psi_{K',\sigma}\longrightarrow
\nonumber \\
&\ & M\Psi^\dagger_{K',\sigma}\sigma_x \exp\Big\{-i\frac{\pi}{2}\sigma_z\Big\} \sigma_z \exp\Big\{i\frac{\pi}{2}\sigma_z\Big\}\sigma_x\Psi_{K',\sigma}= -M \Psi^\dagger_{K',\sigma} \sigma_z \Psi_{K',\sigma}.
\label{1285xy}
\end{eqnarray}
The terms that are linear in the momentum transform in such a way that the $p_x$ term changes sign, whereas the $p_y$ term remains invariant:
\begin{eqnarray}
&\ &\Psi^\dagger_{K',\sigma} \left[-\sigma_x p_x +\sigma_y p_y\right]\Psi_{K',\sigma}\longrightarrow
\nonumber \\
&\ & \Psi^\dagger_{K',\sigma}\sigma_x \exp\Big\{-i\frac{\pi}{2}\sigma_z\Big\} \left[-\sigma_x p_x +\sigma_y p_y\right]\exp\Big\{i\frac{\pi}{2}\sigma_z\Big\}\sigma_x\Psi_{K',\sigma}=
 \Psi^\dagger_{K',\sigma} \left[\sigma_x p_x +\sigma_y p_y\right]\Psi_{K',\sigma}. 
\label{1285xy1}
\end{eqnarray}
Hence we obtain, in the massive case
\begin{eqnarray}
h_K &=& v_F \left[p_x\sigma_x + p_y \sigma_y + M\sigma_z\right ] 
\label{1287xx1}
\end{eqnarray}
and
\begin{eqnarray}
h_{K'} &=& v_F \left[p_x\sigma_x + p_y \sigma_y - M\sigma_z\right ], 
\label{1287xx2}
\end{eqnarray}
where $v_F=\sqrt{3} \tilde{t}a / 2 $.
Notice that the mass at valley $K'$ has an opposite sign to that at valley $K$.  

Now, there is an important issue concerning the Dirac description. The energy eigenvalues of the Dirac Hamiltonians $h_{K}$ and $h_{K'}$ are
unbounded from below: $\varepsilon=\pm\sqrt{|\textbf{p}|^2+M^2}$. In order to solve this problem, Dirac introduced the concept of a ``Dirac sea''.
This is realized in solids with a completely filled valence band, which is precisely the case in TMD's. From the Dirac-sea concept, it follows naturally the concept of a hole, which is the lack of an electron in the Dirac sea.

We now introduce a fully quantized Dirac-field operator $\psi_{Q,\sigma}(\textbf{x},t)$, with $Q=K,K'$, and $\sigma=\uparrow, \downarrow$. 
The corresponding Hamiltonian reads
\begin{equation}
\mathcal{H} =-i v_F \sum_{Q=K,K';\sigma=\uparrow. \downarrow} \psi_{Q,\sigma}^\dagger \sigma^i \nabla_i \psi_{Q,\sigma}, 
\end{equation}
where it is implicitly assumed that the ground state is the Dirac sea or a completely filled valence band, and the associated field equation reads
\begin{equation}
\left[\sigma_i\nabla_i -\beta\partial_t -M\right]\psi_{Q,\sigma}(\textbf{x},t)=0,
\end{equation}
where $\sigma_i$ are the $2\times 2$ Pauli matrices, $i=x,y$ and $\beta =\sigma_z$.

The corresponding solution for the Dirac field is given by
\begin{equation}
\psi_{Q,\sigma}(\textbf{x},t) = \int d^2 p \frac{M}{\omega(\textbf{p})}\left[ c_{Q,\sigma}(\textbf{p})e^{i[\textbf{p}\cdot\textbf{x} - \omega(\textbf{p})t]}u(\textbf{p}) + d^\dagger_{Q,\sigma}(\textbf{p})e^{i[-\textbf{p}\cdot\textbf{x}+\omega(\textbf{p})t]}v(\textbf{p})\right ], 
\end{equation}
where $c_{Q,\sigma}(\textbf{p}) $ is the electron annihilation operator and $ d^\dagger_{Q,\sigma}(\textbf{p})$
is the hole creation operator, with momentum $\textbf{p}$. Both electron and hole possess positive energy $\varepsilon = \hbar \omega(\textbf{p})=\sqrt{|\textbf{p}|^2+M^2}$ (recall that we use units $\hbar=c=1$). The two-component spinors  $u(\textbf{p})$ and $v(\textbf{p})$ satisfy
\begin{equation}
\left[\textbf{p}^i \sigma^i +\beta\omega(\textbf{p}) +M \right]u(\textbf{p})=0, \qquad 
\left[-\textbf{p}^i \sigma^i -\beta\omega(\textbf{p}) +M \right]v(\textbf{p})=0
\label{equv}
\end{equation}
 Notice that contrarily to the field in the tight-binding Eq.(\ref{1330}), which describes electrons, now both electrons and holes are described within the Dirac-field approach. 

So far the electromagnetic interaction has been negected. A full description thereof is achieved by the minimal coupling to the electromagnetic field, which is written in Eq. (1) of the main paper. \\


{\bf 2. Time-Reversal Invariant Hamiltonian} \\

Here, we show that although a mass term in the Dirac equation breaks the time-reversal (TR) symmetry, when there are two valleys, $K$ and $K'$, this symmetry is restored if they have masses with opposite sign. Indeed, under the TR operation, the Dirac field transforms as (see Eq. (16.31) of Ref. \cite{ef})
$$
\psi(\textbf{r},t)\stackrel{TR}{\longrightarrow} - i \sigma_y \psi(\textbf{r},-t).
$$

Considering the invariance of the spacetime volume $\int d^2r dt$ 
under the overall transformation $(\textbf{r},t) \longrightarrow (-\textbf{r},-t)$, 
we may write the TR operation equivalently as \cite{ef}
\begin{equation}
\psi(\textbf{r},t)\stackrel{TR}{\longrightarrow} - i \sigma_y \psi(-\textbf{r},t).
\end{equation}

The mass term
$
\int d^2r\overline{\psi}(\textbf{r},t)\psi(\textbf{r},t),
$
where $\overline{\psi}\equiv \psi^\dagger\sigma_z$, transforms as
\begin{eqnarray}
\int d^2r\overline{\psi}(\textbf{r},t)\psi(\textbf{r},t)=\int d^2r\psi^\dagger(\textbf{r},t)\sigma_z\psi(\textbf{r},t)
\stackrel{TR}{\longrightarrow} 
\nonumber \\
- \int d^2r\overline{\psi}(-\textbf{r},t)\psi(-\textbf{r},t)=- \int d^2r\overline{\psi}(\textbf{r},t)\psi(\textbf{r},t),
\label{TR}
\end{eqnarray}
where we used the fact that $\sigma_y\sigma_z\sigma_y=-\sigma_z$, as well as the invariance of $\int_{-\infty}^\infty \int_{-\infty}^\infty
dxdy$ under $(\textbf{r},t)\longrightarrow (-\textbf{r},t)$.

Let us consider now the flavors: $K, K', \uparrow,\downarrow$. We then have, under the TR operation
$$
\psi(\textbf{r},t)_{K,\uparrow}\stackrel{TR}{\longrightarrow} - i \sigma_y \psi(-\textbf{r},t)_{K',\downarrow}; 
$$
hence
\begin{eqnarray}
\int d^2r\sum_\sigma\overline{\psi}_{K,\sigma}(\textbf{r},t)\psi_{K,\sigma}(\textbf{r},t) \stackrel{TR}{\longrightarrow }
\nonumber \\
- \int d^2r\sum_\sigma\overline{\psi}_{K',\sigma}(\textbf{r},t)\psi_{K',\sigma}(\textbf{r},t).
\label{TRa}
\end{eqnarray}

A TR invariant massive theory, therefore, is obtained when the masses associated to the $K$ and $K'$ valley flavors have opposite signs,
whereas the masses associated to the $ \uparrow,\downarrow$ spin flavors have the same sign. 
In this case, clearly, the sum of mass terms becomes TR invariant:
\begin{eqnarray}
M \int d^2r\sum_\sigma\overline{\psi}_{K,\sigma}\psi_{K,\sigma} - M \int d^2r\sum_\sigma\overline{\psi}_{K',\sigma}\psi_{K',\sigma} \stackrel{TR}{\longrightarrow}
\nonumber \\
-M \int d^2\sum_\sigma\overline{\psi}_{K',\sigma}\psi_{K',\sigma} + M \int d^2r\sum_\sigma\overline{\psi}_{K,\sigma}\psi_{K,\sigma}.
\label{TR1}
\end{eqnarray}
Notice that the spins also flipped, but since we are summing over them, we simply got both terms in inverted order in the addition. 
This is precisely the situation that one finds by Taylor expanding the tight-binding energy around the $K$ and $K'$ valley flavors.
We conclude, therefore, that the gapped Dirac Hamiltonian that we are using for describing TMD's is TR invariant. \\

{\bf 3. The  Foldy-Wouthuysen Expansion: Effects of Spin-Orbit Coupling} \\

Let us consider here the emergence of the spin-orbit coupling out of the quantum electrodynamics minimal coupling. For this purpose, we shall examine the so called Foldy-Wouthuysen expansion of the Dirac Hamiltonian around the non-relativistic limit \cite{fw,iz}. We will concentrate on one of the valleys, say, the $K$ valley.
The free Dirac Hamiltonian in this case is given in momentum space by Eq.~(\ref{1287xx1}) of the manuscript, and admits the following spinor solutions for positive and negative energies, respectively
\begin{eqnarray}
u(p) &=&
\begin{pmatrix}
u_L(p)\\

u_S(p)
\nonumber\\
\end{pmatrix} 
\ \ \ ; \ \ \ 
v(p) =
\begin{pmatrix}
v_L(p)\\

v_S(p)
\end{pmatrix},
\end{eqnarray}
\n where
$$
u_{L}(p)=\sqrt{\frac{\varepsilon+M}{2M}}, \quad
u_S(p)=\sqrt{\frac{\varepsilon+M}{2M}}\frac{{\bf \sigma}\cdot {\bf p}}{\varepsilon+M}=\frac{{\bf \sigma}\cdot {\bf p}}{\varepsilon+M}\,u_L(p), \quad
v_S(p)=-\frac{{\bf \sigma}\cdot {\bf p}}{\varepsilon+M}\,v_L(p), \quad
v_L(p)=\sqrt{\frac{\varepsilon+M}{2M}}.
$$

The non-relativistic limit of the Hamiltonian Dirac equation can be consistently investigated, in any order in $|{\bold p}|/M$, using the Foldy-Wouthuysen transformation \cite{fw,iz}. Such a unitary transformation consists of re-writing the Dirac Hamiltonian in order to separate the spinors' large $(L)$ and small $(S)$ components.
Note that the solutions $u_L(p)$ and $u_S(p)$, as well as $v_L(p)$ and $v_S(p)$ are coupled by the term ${\bf \sigma}\cdot {\bf p}/(\varepsilon+M)$. In the non-relativistic limit, when $|{\bold p}|\ll M$, the component $u_L(p)\propto 1$ while $u_S(p)\propto|{\bf p}|/M$; therefore, $u_L(p)\gg u_S(p)$. These are known as the large and the small components of the positive energy Dirac solution. Similarly for $v_L(p)$ and $v_S(p)$. It is common in the literature to call operators that couple the large and small components as "odd operators" (as $\sigma$,$\gamma$,$\gamma^5$,etc...), whereas those that do not do this coupling are called "even operators"(as $I$ and $\beta\equiv\sigma^z$).  

Specifically, one introduces an unitary transformation $U(\theta)$, such that $\psi^{\prime}=U(\theta)\psi$, and
\begin{equation}
i\partial_t \psi^{\prime}= \mathcal{H}^{\prime} \psi^{\prime}. 
\end{equation}
Inserting the previous expression for $\psi^{\prime}$, we immediately obtain
\begin{equation}
\mathcal{H}^{\prime}=U(\theta)\left[\mathcal{H} -i\partial_t  \right] U^{\dagger}(\theta).
\end{equation}

In the free case, $U(\theta)$ can be written as
\begin{equation}
U(\theta)=\exp\PC{\beta\frac{{\bf \sigma}\cdot {\bf p}}{|{\bf p}|}\theta},
\end{equation}
where $\beta=\sigma^z$. Then, using the properties of the Pauli matrices, we obtain
\begin{eqnarray}
U(\theta)=\cos\theta+\beta\frac{{\bf \sigma}\cdot {\bf p}}{|{\bf p}|}\,\sin\theta.
\end{eqnarray}

It follows that  $U(\theta)U^{\dagger}(\theta)=\mathbb{I}$, which means the transformation is unitary. Under such transformation, the free Dirac Hamiltonian will transform as
\begin{eqnarray}
\mathcal{H}\rightarrow \mathcal{H}^{\prime}&=&U(\theta)\mathcal{H}U^{\dagger}(\theta) \\
&=& {\bf \sigma}\cdot {\bf p}\left(\cos 2\theta-\frac{M}{|{\bf p}|}\,\sin 2\theta\right)+\beta \left( M\cos 2\theta+|{\bf p}|\sin 2\theta \right).
\nonumber
\end{eqnarray}

By choosing $\theta$ such that $\sin 2\theta =|{\bf p}|/ \varepsilon$ and $\cos 2\theta=M/\varepsilon$, the odd operator ${\bf \sigma}\cdot {\bf p}$ is eliminated and we find
\begin{equation}
\mathcal{H}^{\prime}=\beta\sqrt{{\bf p}^2+M^2}.
\end{equation}

For the interacting case, obtained from the minimal coupling $p_{\mu}\rightarrow p_{\mu}-e A_{\mu}$, where $e$ denotes the charge of the particle and $A_{\mu}$ represents the four-vector potential of the associated electromagnetic field, we have to expand $\mathcal{H}^{\prime}$ to higher order by using the well-known Baker-Campbell-Hausdorff formula because the first unitary transformation does not eliminate the odd operators. The procedure described above is repeated until these terms are eliminated from the Hamiltonian, which occurs after the third transformation. The resulting Hamiltonian is then given by \cite{iz,rw}
\begin{equation}
\mathcal{H}^{\prime\prime\prime}= \beta\left[M+\frac{({\bf p}-e {\bf A})^2}{2M}-\frac{{\bf p}^4}{8M^3}\right]+eA^0-\frac{e}{2M}\beta {\bf \sigma}\cdot{\bf B}+\left(-i\frac{e}{8M^2}{\bf \sigma}\cdot{\bf \nabla}\times {\bf E}-\frac{e}{4M^2}{\bf \sigma}\cdot {\bf E}\times {\bf p}\right)-\frac{e}{8M^2}{\bf \nabla}\cdot {\bf E}. 
\label{FW}
\end{equation}
The first term (the sum of terms inside the square brackets) of Eq.~(\ref{FW}) is the expansion of $(({\bf P}-e {\bf A})^2+M^2)^{1/2}$. The second term represents the electrostatic potential of a point charge, whereas the third term represents the Zeeman energy. The fourth term (within parenthesis) is the Pauli spin-orbit coupling, and the last one, represents the Darwin term.

A procedure analogous to the Foldy-Wouthuysen expansion can be made by considering the decoupling of electron (or hole) states belonging to the valence and conduction bands \cite{rw}, in the same way as the Foldy-Wouthuysen expansion decouples large and small components. Starting from the $8 \times 8 $ Kane Hamiltonian \cite{kh}, one arrives at an effective $2 \times 2 $ conduction band Hamiltonian, which coincides with the one obtained by employing the Foldy-Wouthuysen expansion on the Dirac Hamiltonian in the presence of an electromagnetic field (\ref{FW}), except for the fact that the coefficients of each of the terms are now determined by parameters that are directly associated to the conduction and valence bands \cite{rw}. The intensity of each term, consequently, may vary, depending on the values of such parameters.
The Rashba spin-orbit coupling, for instance, which corresponds in this framework to the Pauli spin-orbit term, may be absent for certain values of the band parameters, while in the original Dirac Hamiltonian it is always present because it is an intrinsic feature.

We conclude, therefore, that the spin-orbit coupling is in general automatically included within the description of the electromagnetic interactions obtained by the minimal coupling of the electromagnetic field with the Dirac field. Moreover, the parabolic band emerges naturally out of the hyperbolic massive Dirac dispersion relation.
 \\

{\bf  4. Bethe-Salpeter Equation and Exciton States} \\

The exciton propagator is given by the two-point correlation function of the exciton creation operator $\varphi(x)$, namely 
\begin{equation}
G(x;y)= \langle 0|T \varphi(x) \varphi(y)^\dagger|0\rangle,  
\end{equation}
where $T$ is the time-ordering operator.
Since the exciton is an electron-hole bound state, it follows that the exciton creation operator is composed of an electron and a hole creation operators, namely $\varphi(x) =\psi(x) \psi^\dagger(x)$. Consequently, the exciton propagator is given by the four-point function, in the following limit:

\begin{equation}
G(x;y)=\lim_{x_1\rightarrow y_1;x_2\rightarrow y_2}G(x_1,y_1;x_2,y_2)= \lim_{x_1\rightarrow y_1;x_2\rightarrow y_2}\langle 0|T \psi(x_1) \psi^\dagger(y_1) \psi(x_2) \psi^\dagger(y_2)|0\rangle, 
\end{equation}
which has the Fourier transform
$\lim_{p_1,p_2\rightarrow 0}G(p_1, p_2,k) = G(k).$

By using the perturbation expansion on the interaction Hamiltonian $\mathcal{H}_{int}= e \overline \psi \gamma^\mu \psi A_\mu$, which describes the full electromagnetic interaction, one can calculate the above four-point function up to the desired order.
Nevertheless, it shall prove more convenient to follow a different way. Indeed,
$G(p_1, p_2,k)$ obeys the Bethe-Salpeter equation \cite{iz}
\begin{eqnarray}
S^{-1}(p_1) S^{-1}(p_2) G(p_1,p_2,k) = \mathbb{I} +\int \frac{d^3q}{(2\pi)^3}\Gamma (q)  G(p_1,p_2,k-q),
\label{bs} 
\end{eqnarray}
where $\Gamma (q)$ is the interaction kernel and
\begin{equation}
 S(p)=\frac{1}{p\!\!\!/ - M - \Sigma(p)},
\end{equation}
is the exact electron-hole propagator.

The Bethe-Salpeter equation is formally written as
\begin{eqnarray}
S^{-1} S^{-1} G &=& 1 + \Gamma  G
\nonumber \\
G &=& SS + SS \Gamma G
\nonumber \\
\left[ 1 - SS \Gamma \right ] G &=& SS,
\label{bs1} 
\end{eqnarray}
and it is solved by
\begin{eqnarray}
 G &=& \frac{1}{1 - SS \Gamma  } SS.
\label{bs2} 
\end{eqnarray}

The inverse exciton propagator, therefore, can be written as
$
G^{-1}= S^{-1} S^{-1} - \Gamma.
$
Since the free-exciton propagator is just $G_0 = S^2$, the above equation may be re-written as
\begin{equation}
G^{-1}(p)=G_0^{-1}(p) - \Gamma(p).
\end{equation}

Hence, the exciton propagator $G(p)$ itself satisfies a Schwinger-Dyson equation, with the interaction kernel $\Gamma(p)$ and $S^2$ playing the role, respectively, of the self-energy and of the free-exciton propagator.

From the above equation, we obtain
\begin{equation}
	G(p)=\frac{1}{\Big[p\!\!\!/ - M - \Sigma(p) \Big]^2- \Gamma}.
\end{equation}

In the next section, we show that the exciton-mass spectrum  derived from the inner electron-hole structure is given by the renormalized mass $M_R=M+\Sigma(p\!\!\!/=M_R)$. The $\Gamma$ kernel provides the mass renormalization due to the exciton-exciton interaction.  We are going to neglect the $\Gamma$-kernel contribution to the mass renormalization, thus focusing just on the exciton inner structure, independently of how the excitons interact among themselves. We thereby obtain the exciton mass spectrum from $M_R=M+\Sigma(p\!\!\!/=M_R)$, where $\Sigma(p\!\!\!/=M_R)$ will be determined from the Schwinger-Dyson formalism. 

Figure 1, representing the electron and hole world-lines, conveys the idea behind our approximation for solving the Bethe-Salpeter equation.

\begin{figure}[h]
\centering
\includegraphics[scale=0.50]{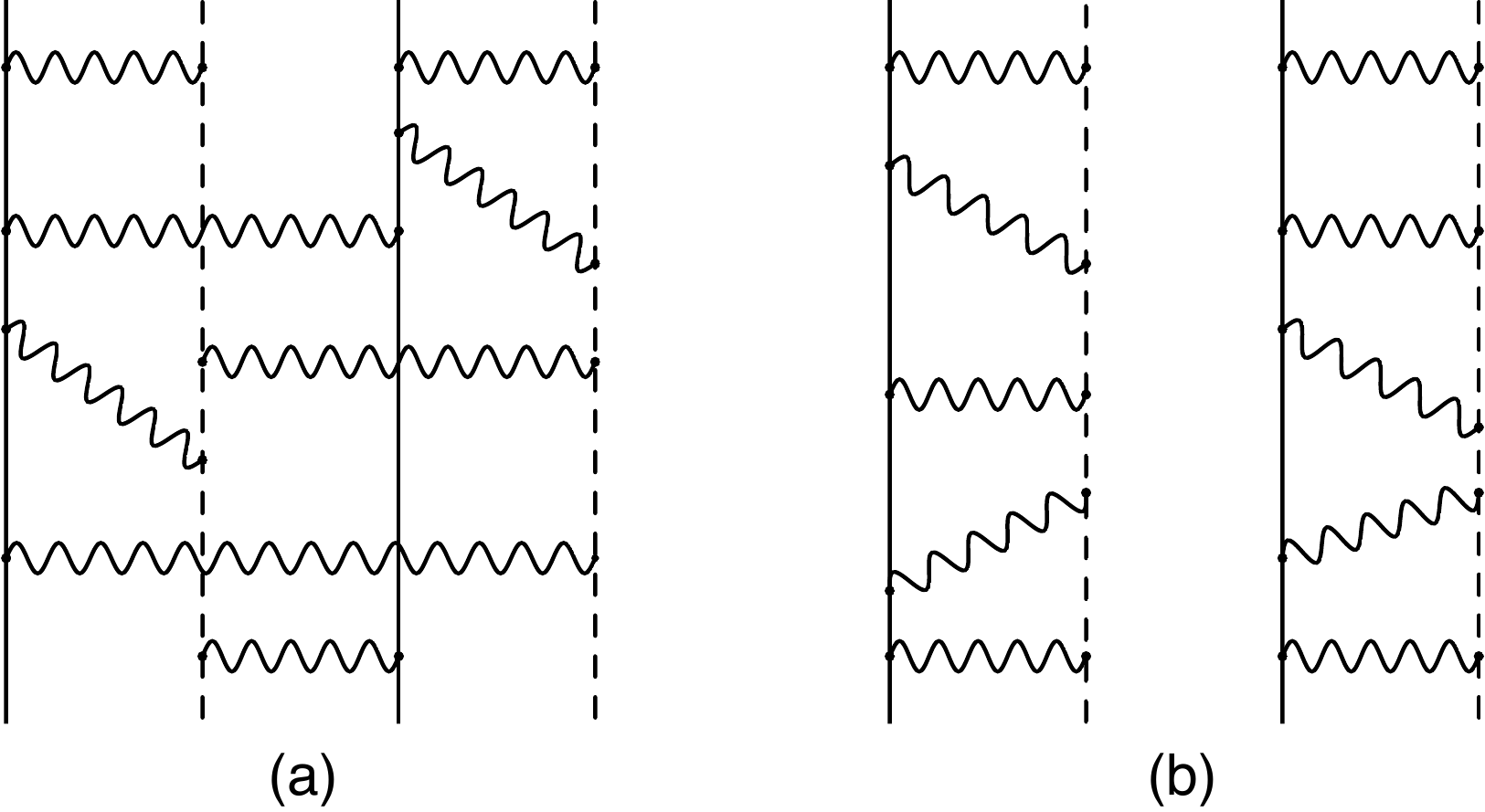}
\caption{Schematic representation of our approximation for solving the Bethe-Salpeter equation. Solid and broken lines represent, respectively, the world-lines of electrons and holes. Notice that these are not Feynman diagrams. a) The complete treatment of the problem would include interactions among all electrons and holes. b) In our approximation, we just consider the interactions between a given electron and its associated hole; the two of which will form an exciton bound-state. Different excitons, therefore, do not interact within this approximation.} \label{Fig1}
\end{figure}

\begin{figure}[h]
\centering
\includegraphics[scale=0.4]{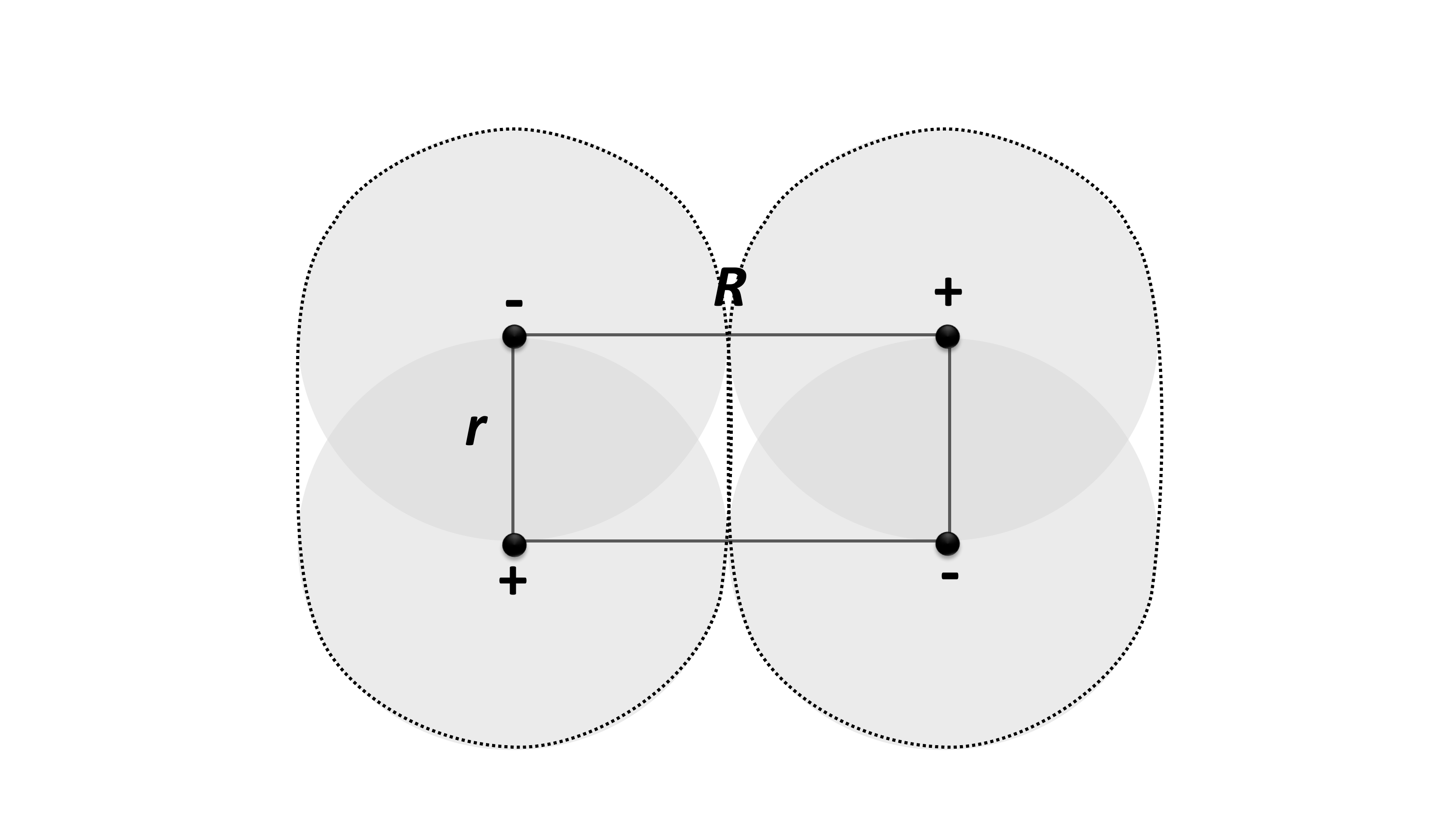}
\caption{Sketch of a bounded exciton, where each electron-hole pair is delimitated by the dotted black lines. $R$ is the distance between two excitons, and $r$ is a characteristic distance between positive and negative charges to form excitons.} \label{Fig2}
\end{figure}

Below, we provide a rough estimate of the exciton-exciton interaction energy $E_2$, and show that it is approximately one order of magnitude smaller than the exciton binding energy $E_1$, thus justifying our approximation that neglects $\Gamma$.
Let us consider four elementary charges, two positive and two negative, disposed at the vertices of a rectangle with sides $r$ and $R>r$, as shown in Fig.~\ref{Fig2}. Excitons are formed between the positive and negative charges separated by the small distance $r$, and each exciton has the energy $E_1=-e^2/r$. The electrostatic energy $E_2$ between the two excitons is then

\begin{equation}
E_2=-2e^2\left[\frac{1}{R}-\frac{1}{\sqrt{r^2+R^2}}\right],
\end{equation}
where $R$ is the distance between the pairs. In this case, the ratio between the inter-exciton and the intra-exciton interactions read
\begin{equation}
\frac{E_2}{2E_1}\approx \frac{1}{2}\PC{\frac{r}{R}}^{3}, \label{E2E1}
\end{equation}
where we used $R>>r$. Since the minimal distance $R \approx 2r$ (see Fig.~\ref{Fig2}), Eq.~(\ref{E2E1}) yields $E_2/2E_1\approx 1/16$. Our rough estimate indicates that the exciton-exciton interaction is at least approximately 15 times smaller than the electron-hole excitations. This is corroborated by the fact that the excitons studied here have a binding energy of the order of 300 meV, whereas the bi-excitons found in such materials have a reported binding energy of the order of 20 meV \cite{Hao2017}. Our approximation is thus well justified.
\\

{\bf 5. Schwinger-Dyson Equation} \\

Next, we provide details about how the Schwinger-Dyson method can be used for obtaining the dynamical generation of midgap exciton bound states. Here, we neglect the quantum corrections to the interacting vertex function. The full electron propagator reads
\begin{equation}
S_F^{-1}(p)=S_{0F}^{-1}(p)-\Sigma(p),  \label{fullpropfer2cap5}
\end{equation}
where the electron self-energy $\Sigma(p)$ is given by
\begin{eqnarray}\label{fermioncap5}
\Sigma (p)=e^2\int\frac{d^3k}{(2\pi)^3} \gamma^{\mu}S_F(k) \gamma^{\nu}\,G_{\mu\nu}(p-k),
\end{eqnarray}
and $G_{\mu\nu}$ is the gauge-field propagator. To calculate the poles of the full-electron propagator, we first perform a Taylor expansion in the electron self-energy in Eq.~(\ref{fermioncap5}) around $M^R_a$, yielding 
\begin{eqnarray}
\Sigma(p)=\Sigma(p\!\!\!/=M^R_a)+(\gamma^\mu p_\mu-M^R_a)\frac{\partial\,\Sigma(p)}{\partial p\!\!\!/}\huge|_{p\!\!\!/=M^R_a}+...\end{eqnarray}

The physical energy spectrum may be obtained by imposing 
\begin{equation}
\Sigma (p\!\!\!/=M^R_a)=M^R_a -M_a, \label{cond1}
\end{equation}
where the bare mass $M_a$ is given by
\begin{equation}
M_{a,\sigma}=(M_{K}, M_{K'})_{\sigma}= (\Delta,-\Delta)_{\sigma}.  \label{baremass}
\end{equation}

The full fermion propagator in Eq.~(\ref{fullpropfer2cap5}) may be written as
\begin{eqnarray}
S_F(p)&=&\frac{1}{\gamma^\mu p_\mu-M_a-\Sigma(p)}\nonumber\\
&=&\frac{1}{(\gamma^\mu p_\mu-M^R_a)(1-\frac{\partial\,\Sigma(p)}{\partial p\!\!\!/}\huge|_{p\!\!\!/=M^R_a}+...)}
\nonumber\\
&=&\frac{\gamma^\mu p_\mu+M^R_a}{(p^2 - M{^R_a}^2)(1-\frac{\partial\,\Sigma(p)}{\partial p\!\!\!/}\huge|_{p\!\!\!/=M^R_a}+...)}.
\label{sfsup}
\end{eqnarray}

We find that  $M^R_a$ is the pole of the full physical electron propagator at zero momentum, being therefore
the desired physical mass. This result is exact. Nevertheless, in order to obtain analytical solutions of the Schwinger-Dyson equation, it is convenient to rewrite the full fermion propagator as \cite{Maris2}
\begin{equation}
S_F^{-1}(p)=p_{\mu}\gamma^{\mu}\,A(p)+\Sigma_a (p), \label{full1}
\end{equation}
where $A(p)$ is the usually called wavefunction renormalization and $\Sigma_a(p)$ the mass function. From Eq.~(\ref{full1}) and Eq.~(\ref{fullpropfer2cap5}), we obtain $\Sigma(p)=-\Sigma_a(p)-M_a$. Therefore,
\begin{equation}
\Sigma_a(p\!\!\!/=\mp |M{^R_a}|)=\mp |M^R_a|, \label{polefinal}
\end{equation}
is the equation that we have to solve to find the physical mass $M^R_a$. We then replace Eq.~(\ref{full1}) into Eq.~(\ref{fullpropfer2cap5}), and calculate the trace over the Dirac matrices. After some standard operations, we find that the integral equation for $\Sigma_a(p)$ is
\begin{equation}
\Sigma_a(p)=-M_a-4\pi \alpha\int \frac{d^3 k}{(2\pi)^3}\frac{\Sigma_a(k)}{A^2(k) k^2+\Sigma_a^2(k)} \delta^{\mu\nu}G_{\mu\nu}(p-k). \label{eqintsiga1}
\end{equation}

Eq.~(\ref{eqintsiga1})  does not have an analytical solution; hence, from now on we shall make approximations that are common in the literature. The first one concerns the gauge-field propagator $G_{\mu\nu}$,
\begin{equation}
G^{-1}_{\mu\nu}=G^{-1}_{0,\mu\nu}-\Pi_{\mu\nu}, \label{SDphoton}
\end{equation}
or equivalently,
\begin{equation}
G_{\mu\nu}=G_{0,\mu\alpha}[\delta^{\alpha}_\nu-\Pi^{\alpha\beta}G_{0,\beta\nu}]^{-1},\label{solphoton}
\end{equation}
where $G_{\mu\nu}$ is the full propagator of the gauge field and $\Pi_{\mu\nu}$ is the vacuum polarization tensor, which may be decomposed into
\begin{equation}
\Pi_{\mu\nu}=e^2 \Pi_1 P_{\mu\nu}+e^2\Pi_2\epsilon_{\mu\nu\alpha}p^\alpha.  \label{Pidec}
\end{equation}

Using $G_{0,\mu\nu}=P_{\mu\nu}/2 p$, with $P_{\mu\nu}=\delta_{\mu\nu}-p_\mu p_\nu/p^2$ and $p\equiv \sqrt{p^{2}}$, and inserting Eq.~(\ref{Pidec}) into Eq.~(\ref{solphoton}), we find
\begin{equation}
G_{\mu\nu}(p)=\Delta_1(p) P_{\mu\nu}+ \Delta_2(p) \epsilon_{\mu\nu\alpha}p^\alpha,  \label{exacpho}
\end{equation}
where
\begin{equation}
\Delta_1(p)=\frac{2 p-e^2 \Pi_1}{(2p-e^2 \Pi_1)^2+e^4p^2 \Pi^2_2}
\end{equation}
and
\begin{equation}
\Delta_2(p)=\frac{e^2\Pi_2}{(2p-e^2 \Pi_1)^2+e^4p^2 \Pi^2_2}.
\end{equation}

We immediately see that only the first term of Eq.~(\ref{Pidec}) contributes to Eq.~(\ref{eqintsiga1}). Consequently, we use
only the first term of the propagator, namely,
\begin{equation}
G_{\mu\nu}(p) \rightarrow \Delta_1(p) P_{\mu\nu}.  \label{apppho}
\end{equation}
On the other hand, using the results in Ref.~\cite{Redlich}, it follows that, for large momentum, $\Pi_1\rightarrow p$ and $\Pi_2\rightarrow 1/p$. Hence, we may write
\begin{equation}
\Delta_1(p)\approx \frac{1}{p} \left[ -C_1(\lambda)+O\left(\frac{M_a}{p}\right) \right], \label{Delta1S46}
\end{equation}
with
\begin{equation}
C_1(\lambda)=\frac{16(\lambda-32)}{1024-\lambda[64-\lambda]}, \label{C1}
\end{equation}
and $\lambda=e^2 N_f=4\pi\alpha N_f$.

Inserting Eq.~(\ref{Delta1S46}) in Eq.~(\ref{eqintsiga1}) written in spherical coordinates, we obtain
\begin{equation}
\Sigma_a (p)=-M_a+8\pi\alpha\int_0^\infty \int_0^\pi\frac{\sin \theta d\theta k^2dk}{(2\pi)^2}\,\frac{\Sigma_a(k)\,}{A^2(k)k^2+\Sigma_a^2(k)}\,\left(\frac{C_1(\lambda)}{\sqrt{q^2}}
\right),
\end{equation}
where $\sqrt{q^2}=|p-k|=\sqrt{p^2+k^2-2pk\cos\theta}$. By defining $u\equiv p^2+k^2-2pk \cos\theta$ and changing the integral variable $\theta$ into $u$, we find after performing the integral over $u$ that
\begin{equation}
\Sigma_a (p)=-M_a+\frac{2\alpha}{\pi p}\int_0^\infty \frac{k dk \Sigma_a(k)}{A^2(k)k^2+\Sigma_a^2(k)}\left\{C_1(\lambda)[|p+k|-|p-k|]
\right\}. \label{c2}
\end{equation}

Now, we introduce a momentum cutoff $\Lambda$ in Eq.~(\ref{c2}), and linearize the logarithmic kernel, using
\begin{equation}
\ln\left[\frac{p+k}{|p-k|}\right]\approx \frac{2k}{p}\Theta(p-k)+\frac{2p}{k}\Theta(k-p),
\end{equation}
to find
\begin{eqnarray}
\Sigma_a (p)=-M_a&+&\frac{2\alpha C_1(\lambda)}{\pi}\left\{ \int_0^p \frac{k^2 dk \Sigma_a(k)}{A^2(k)k^2+\Sigma_a^2(k)} \frac{2}{p} + \int_p^\Lambda \frac{k^2 dk \Sigma_a(k)}{A^2(k)k^2+\Sigma_a^2(k)} \frac{2}{k} \right\}. \label{c3}
\end{eqnarray}

Notice that the bare mass $M_a$ is unphysical, and always smaller than the momentum $p$. The large gap in the problem is actually the ultraviolet energy-momentum cutoff $\Lambda$, and not $\Delta$. 
The next step is to convert Eq.~(\ref{c3}) into a differential equation. In order to do so, we must calculate its derivatives. Note that the derivative of an arbitrary function $F(p)$, given by
\begin{equation}
F(p)=\int_{g(p)}^{h(p)} dk f(k,p),
\end{equation}
is
\begin{equation}
\frac{dF(p)}{dp}=\int_{g(p)}^{h(p)} dk \frac{\partial f(k,p)}{\partial p}+ \frac{\partial h}{\partial p} f\textbf{(}p, h(p)\textbf{)}-\frac{\partial g}{\partial p} f\textbf{(}p, g(p)\textbf{)}, \label{c5}
\end{equation}
where $h(p)$ and $g(p)$ are also arbitrary functions. Therefore, 
\begin{equation}
\frac{d\Sigma_a (p)}{dp}=\frac{d}{dp}\left[\frac{1}{p}\left(C_1(\lambda)\right)\right] \frac{4\alpha}{\pi}\int_0^p \frac{k^2 dk \Sigma_a(k)}{A^2(k)k^2+\Sigma_a^2(k)}. \label{c6}
\end{equation}

By taking the derivative again, we obtain
\begin{equation}
\frac{d}{dp}\left(p^2\frac{d\Sigma_a(p)}{dp}\right)+\frac{4\alpha C_1(\lambda)}{\pi}\frac{p^2\Sigma_a(p)}{A^2(p)p^2+\Sigma_a^2(p)}=0. \label{c7}
\end{equation}

Eq.~(\ref{c7}) is a complicated nonlinear differential equation in $p$. However, it is possible to neglect all the nonlinear behavior using $A^2(p)p^2+\Sigma_a^2(p)\approx p^2$. Indeed, because $A(p)=1+O(\alpha)$ and $\Sigma_a(p)\propto O(\alpha)$, we find 
\begin{equation}
\frac{d}{dp}\left(p^2\frac{d\Sigma_a(p)}{dp}\right)+\frac{4\alpha \,C_1(\alpha)}{\pi}\Sigma_a(p)=0, \label{Euler}
\end{equation}
where $C_1(\alpha)$ is given by Eq.~(\ref{C1}), after replacing $\lambda=4\pi\alpha N_f$. It has been shown in Ref.~\cite{VLWJF} that this linearized regime captures the critical behavior in Eq.~(\ref{c3}). 

The solutions of Euler's differential equation are
\begin{equation}
\Sigma_a(p)={\tilde C} p^{a_{+}}+ {\tilde D} p^{a_{-}},
\label{EulerSol}
\end{equation}
where $a_{\pm}=-1/2\pm i\gamma$, with
the parameter $\gamma$ given by
\begin{eqnarray}
\gamma = \frac{1}{2}\sqrt{\frac{16 \alpha}{ \pi(\pi \alpha - 2) }- 1},
\label{1325} 
\end{eqnarray}
while ${\tilde C}$ and ${\tilde D}$ are constants.

We may rewrite Eq.~(\ref{EulerSol}) as
\begin{eqnarray}
\Sigma_a = \frac{A_{\Lambda}}{\sqrt{p}} e^{-i\left[\gamma \ln\frac{p}{\Lambda}+ \varphi \right]},
\label{1324} 
\end{eqnarray}
where $A_{\Lambda}$ and $\varphi$ are arbitrary real constants. Note that $\gamma$ is real for $\alpha>\alpha_c=2/\pi \approx 0.63$.

Using Eq.~(\ref{polefinal}), choosing $A_{\Lambda}=\Lambda^{3/2}$ (by dimensional reasons) and $\varphi=0$, we get
\begin{eqnarray}
\mathrm{Re}\ \Sigma_a= \frac{\Lambda^{3/2}}{\sqrt{M^R_a}} \cos\left[\gamma \ln\frac{M^R_a}{\Lambda} \right]= M^R_a,
\label{1326x1} 
\end{eqnarray}
and
\begin{eqnarray}
\mathrm{Im}\ \Sigma_a= \frac{\Lambda^{3/2}}{\sqrt{M^R_a}} \sin\left[\gamma \ln\frac{M^R_a}{\Lambda^2} \right].
\label{1326x2} 
\end{eqnarray}

Now, because of Eq.~(\ref{1324}), however, the masses of each valley have different signs, thus leading to two classes of energy eigenstates, according to Eq.~(\ref{polefinal}). The first class of eigenenergies  correspond to a positive mass. In this case, we choose $A_{\Lambda}=\Lambda^{3/2}$ and $\varphi=0$, as we did before. Using Eq.~(\ref{1324}), we conclude that the  eigenstates of the Hamiltonian occur at
\begin{eqnarray}
\varepsilon_+ =\pm \sqrt{|\textbf{p}|^2 + (M^{+}_R)^2},
\label{1350} 
\end{eqnarray}
with
\begin{eqnarray}
M^+_n =  \Lambda e^{-X_n/\gamma}\ \ \ ;\ \ \ n=0,1,2,...
\label{1351} 
\end{eqnarray}
where the $X_n$ are the solutions of the transcendental equation 
\begin{eqnarray}
 e^{-\frac{3x}{2\gamma}} = \cos x.
\label{1328} 
\end{eqnarray}

A second class of eigenenergies corresponds to a negative mass. Choosing $A_{\Lambda}=\Lambda^{3/2}e^{-\pi\gamma}$ and $\varphi=0$, and again using Eq.~(\ref{1324}), we conclude that now the eigenstates of the Hamiltonian occur at
\begin{eqnarray}
\varepsilon_- =\pm \sqrt{|\textbf{p}|^2 + (M^{-}_R)^2},
\label{1352} 
\end{eqnarray}
with
\begin{eqnarray}
M^-_n = - \Lambda e^{-Y_n/\gamma}\ \ \ ;\ \ \ n=0,1,2,...
\label{1353} 
\end{eqnarray}
where the $Y_n$ are the solutions of the transcendental equation
\begin{eqnarray}
 e^{-\frac{3x}{2\gamma}} =-  \sin x.
\label{1354} 
\end{eqnarray}

Observe that $X_0=0$ is a solution with $M>0$, hence a natural choice in the present case is  $\Lambda= |M|$. With this choice, we see that all the states generated by the interaction are located inside the energy gap $[-|M|,+|M|]$. This strongly suggests that they are bound-states of quasi-particles and quasi-holes. \\

{\bf 6. Comparison with Experiment: Exciton Energies} \\

For the sake of conciseness, we describe below in detail the calculations for WS$_2$. A similar procedure was applied for all the other compounds. 

\textit{Monolayer of} WS$_2$: 

In Ref.~\cite{WS2exp}, the authors have measured $n=5$ exciton states for WS$_2$ (see Fig.~3 in Ref.~\cite{WS2exp}). We extract the binding energy (the energy splitting between the bare energy gap and the exciton energy level) for each state, $\varepsilon^{{\rm exp}}_{b,n}=\{-0.3155, -0.1584,-0.0989,-0.0664,-0.0359\}$ eV. Our theoretical results given by Eqs.~(\ref{1351}) and (\ref{1353}) allow to obtain the binding energy for each energy level $\varepsilon^{{\rm Theo.}}_{b,n}$. These are given by alternating the solutions $X_n$ and $Y_n$, i.e., we need to calculate $\{X_1,Y_1,X_2,Y_2,X_3\}$, $\Lambda$, and $\gamma$. Let us first fix the values of $\Lambda$ and $\gamma$ using the first and last experimental points,
\begin{equation}
0.3155=\Lambda \exp\left(-\frac{X_1}{\gamma}\right), \,\, \cos X_1=\exp\left(-\frac{3X_1}{2\gamma}\right)
\end{equation}
and
\begin{equation}
0.0359=\Lambda \exp\left(-\frac{X_3}{\gamma}\right), \,\, \cos X_3=\exp\left(-\frac{3X_3}{2\gamma}\right).
\end{equation}
By solving these four equations in a self-consistent manner, we find $\Lambda=0.4066$ eV and $\gamma=3.2251$, from the fitting for $X_1=0.8183$, and $X_3=7.8277$. Using these results, we obtain $Y_1=3.3533$, $X_2=4.8189$, and $Y_2=6.2279$ from Eqs.~(\ref{1328}) and (\ref{1354}), without any fitting parameter. Therefore, $\varepsilon^{{\rm Theo.}}_{b,n}=\{-0.3155,-0.1437,-0.0912,-0.0589,-0.0359\}$ eV. Because $\gamma=3.2251$, we may use its definition to find $\alpha=0.6618$, which is in excellent agreement with the corresponding value in Ref.~\cite{MarkalphapSupMat}, namely $\alpha=0.7$. \\

{\bf  7. Comparison with Experiment: Exciton Lifetimes} \\

The lifetime of the bound states may be estimated through the ${\rm Im}\ \Sigma_{a}$ given by Eq.~(\ref{1326x2}).
Using Eqs.~(\ref{1351}) and (\ref{1353}), we find for the positive mass solutions
\begin{eqnarray}
\frac{2h}{\tau(X_n)}= \Lambda\frac{\sin X_n}{\left(\cos X_n\right)^{1/3}},
\label{1355} 
\end{eqnarray}
whereas for the negative mass solutions, 
\begin{eqnarray}
\frac{2h}{\tau(Y_n)}= \Lambda\frac{\cos Y_n}{\left(\sin Y_n\right)^{1/3}}.
\label{1356} 
\end{eqnarray}

Using the corresponding transcendental equation for each case, we can express the lifetimes of the midgap states, both for positive and negative masses, respectively, as
\begin{eqnarray}
\tau(X_n) = \frac{2h}{\Lambda}\frac{e^{-{X_n}/2\gamma}}{\sqrt{1- e^{-3{X_n}/\gamma}}}= \frac{2h}{\Lambda}\frac{\left(\cos X_n\right)^{1/3}}{\sin X_n},
\nonumber \\
\nonumber \\
\tau(Y_n)=  \frac{2h}{\Lambda}\frac{e^{-{Y_n}/2\gamma}}{\sqrt{1- e^{-3{Y_n}/\gamma}}}= \frac{2h}{\Lambda}\frac{\left(\sin Y_n\right)^{1/3}}{\cos Y_n}.
\label{1329xyz} 
\end{eqnarray}

Notice that the state $M^+_0$ corresponds to $X_0=0$, and hence has an infinite lifetime, a result that one should expect because according to the choice $\Lambda=|M|$, this state just consists in a free pair of quasiparticle-quasihole with zero binding energy.

Observe also that as $n$ increases, the roots $X_n$ and $Y_n$ rapidly tend to the zeros, respectively, of the cosine and sine functions, and, we therefore conclude that the lifetime of the $M^{\pm}_n$ bound states rapidly tends to zero upon increasing $n$. For the first bound states, we estimate $\tau= 48.34$ fs for  MoS$_2$; $\tau= 32.4$ fs for  WSe$_2$ and $\tau= 24.7$ fs for  WS$_2$. These are direct electron-hole annihilation into a photon.
Now, the photon momentum is very small, because $|\textbf{k}|=\omega/c$ and thus usually cannot accommodate the initial momentum. Then, it follows that exciton decays occur through a more complicated process, in which the exciton lifetime will be considerably enlarged, depending on the temperature. Indeed, it was shown in Ref.\cite{excitlife} that the effective exciton decay time is
\begin{equation}
\tau_n^{eff} = \frac{3k _BT}{2 E_0}\tau_n
\end{equation}
where $E_0 \sim 10^{-5}\ eV$ is the exciton kinetic energy.
At $T=7$ K, this lifetime is enlarged by a factor of about two orders of magnitude, thus yielding values of a few picoseconds for the effective exciton lifetimes, which are the typical values measured in experiments at such temperature \cite{excitlife}:  $\tau_{eff}= 4.2$ ps for  MoS$_2$; $\tau_{eff}= 2.8 $ ps for  WSe$_2$ and $\tau_{eff}= 2.16$ ps for  WS$_2$.\\

{\bf 8. The Keldysh Potential from Pseudo-QED} \\

The simplest approach to 2D excitons in monolayer materials exhibiting a relativistic-like dispersion relation would be to generalize the Wannier-Mott model, by solving the Dirac equation with a Coulomb potential in 2D. This, however, shows a poor agreement with the experimental data, as can be seen in Fig.~3 of Ref.~\cite{WS2exp}. Better results were obtained by the use of the Keldysh two-body potential \cite{KeldyshpSupMat} with an effective dielectric constant $\epsilon_n$ for each bound-state $n$ \cite{WS2exp}. This potential has been derived by Keldysh to describe the electrostatic interaction in thin semiconductors \cite{KeldyshpSupMat},
\begin{equation}
V(r)=\frac{e^2}{8r_0}\left[H_0\left(\frac{r}{r_0}\right)-Y_0\left(\frac{r}{r_0}\right)\right], \label{keldpot}
\end{equation} 
where $r_0$ is a characteristic length, $H_0$ is the Struve function, and $Y_0$ is the Bessel function of second kind. 

Here we shall show that the Keldysh potential emerges from the Pseudo-QED formulation, when we determine the  potential between static charges. The static interaction potential between charged particles in a quantum field theory is given by
\begin{equation}
V(r)=e^2 \int \frac{d^2 \textbf{p}}{(2\pi)^2} \exp(-i \textbf{p}. \textbf{r})  G_{00}(p_0=0,\textbf{p}), \label{potger}
\end{equation}
where $G_{\mu\nu}$ is the corrected mediating gauge-field propagator, given by the Schwinger-Dyson equation for that field, namely
\begin{equation}
G^{-1}_{\mu\nu}=G^{-1}_{0,\mu\nu}-\Pi_{\mu\nu}, \label{sdph}
\end{equation}
where $\Pi_{\mu\nu}$ is the vacuum polarization tensor. The 00-component of $\Pi_{\mu\nu}$ reads
\begin{equation}
\Pi_{00}(p_0=0,\textbf{p})= -\frac{N_f e^2}{2\pi} \int_0^1 dt \, \frac{t(1-t) \textbf{p}^2}{\sqrt{\Delta^2+t(1-t)\textbf{p}^2}},
\equiv \textbf{p}^2 F(\textbf{p}). 
\label{pi00}
\end{equation}
where $N_f$ is the number of flavors.

By substituting Eq.~(\ref{sdph}) and Eq.~(\ref{pi00}) in Eq.~(\ref{potger}), we find the static potential 
\begin{equation}
V(r)=e^2 \int\frac{d^2 \textbf{p}}{(2\pi)^2} \frac{\exp(-i \textbf{p} . \textbf{r})}{2\sqrt{\textbf{p}^2} - \textbf{p}^2F(\textbf{p})}.
\label{vr}
\end{equation}

At short ($r\rightarrow 0$) and large ($r\rightarrow \infty$) distances, the integrand is dominated, respectively, by $\textbf{p}\rightarrow \infty$ and $\textbf{p}\rightarrow 0$.

We analize firstly the large distance (small momentum) regime. In the small-momentum limit, $\Pi_{00}$ reads 
\begin{equation}
\Pi_{00}(p_0=0,\textbf{p})\approx -\frac{N_f e^2}{2\pi} \int^1_0 dt \frac{t(1-t)\textbf{p}^2}{|\Delta|}=-\frac{N_f e^2}{4\pi}\frac{\textbf{p}^2}{3|\Delta|}=-\frac{N_f  \alpha \textbf{p}^2}{3|\Delta|}. \label{pi002}
\end{equation}
Hence, $\lim_{p\rightarrow 0} F(\textbf{p}) = -N_f  \alpha / 3 |\Delta|$. 
Using polar coordinates $d^2 \textbf{p}=p dp d\theta$ with $p=\sqrt{\textbf{p}^2}$ (in static regime), we have 
\begin{equation}
V(r)\approx e^2 \int_0^\infty\frac{pdp}{2\pi}\int_0^{2\pi}\frac{d\theta}{2\pi} \frac{\exp(-i p r \cos\theta)}{2 p+N_f \alpha p^2/3|\Delta|}.
\end{equation}
Integrating $\theta$ out, it is easy to verify that
\begin{equation}
V(r)\approx e^2\int_0^\infty \frac{pdp}{2\pi} \frac{J_0(pr)}{2p+N_f \alpha p^2/3 |\Delta|},
\end{equation}
where $J_0$ is the Bessel function of first kind. Integrating out $p$ and defining $r_0\equiv N_f\alpha/ 6 |\Delta|$, we obtain the large distance behavior of our static potential:
\begin{equation}
V(r) \approx \frac{e^2}{8r_0}\left[H_0\left(\frac{r}{r_0}\right)-Y_0\left(\frac{r}{r_0}\right)\right],\qquad r_0=\frac{N_f \alpha}{6|\Delta|}, \label{pqedpot}
\end{equation}
which is exactly the Keldysh potential given in Eq.~(\ref{keldpot}). In physical units, $r_0=\hbar N_f v_F \alpha/6|\Delta|$. For $r \gg r_0$, we have
\begin{equation}
V(r)\approx \frac{e^2}{4\pi r}\left[1-\frac{r_0^2}{r^2}+O\left(\frac{r^4_0}{r^4}\right)\right],
\end{equation}
which shows that the Keldysh potential yields the usual Coulomb interaction in the lowest order.
Let us consider now the small distance (large momentum) regime in Eq.~(\ref{vr}). In this case, $\lim_{p\rightarrow \infty} F(\textbf{p}) = -N_f  e^2 / 16|\textbf{p}|$, and after inserting this result in Eq. (\ref{vr}), we find that the static potential is proportional to the Fourier transform of $1/ |\textbf{p}|$, which is the familiar $1/r$-Coulomb potential with an overall factor $1/(1+N_f e^2/16)$.

In conclusion, the static potential of PQED, given by Eq.~(\ref{vr}), behaves as the $1/r$ Coulomb potential at short distances $r \ll r_0$ and as the Keldysh potential at large distances $r \gg r_0$, which, on its turn, is also a Coulomb potential in a lowest-order expansion. \\


\begin{thebibliography}{100}
\bibitem{footnote} Notice that in the lowest-order expansion, the Keldysh potential that arises for large distances is itself a Coulomb potential.
\end{thebibliography}
\end{document}